\documentclass{svmult}

\usepackage{makeidx}         
\usepackage{graphicx}        
\usepackage{multicol}        
\usepackage[bottom]{footmisc}

\usepackage{newtxtext}       %
\usepackage{newtxmath}       

\usepackage{url}

\makeindex             

\begin{document}

\title*{Opinion Network Modeling and Experiment}
\author{Michael Gabbay}
\institute{Michael Gabbay \at  Applied Physics Laboratory, University of Washington, 1013 NE 40th St, Seattle, WA, USA 98105-6698, \email{gabbay@uw.edu}. \\
This is a pre-copyedited version of a contribution published in \emph{Proceedings of the 5th International Conference on Applications in Nonlinear Dynamics}, V. In, P. Longhini, \& A. Palacios (eds.) published by Springer International Publishing (2019). The definitive authenticated version is available online via \url{https://doi.org/10.1007/978-3-030-10892-2_18}.}

%
%
\maketitle


\abstract{We present a model describing the temporal evolution of opinions due to interactions among a network of individuals. This Accept-Shift-Constrict (ASC) model is formulated in terms of coupled nonlinear differential equations for opinions and uncertainties. The ASC model dynamics allows for the emergence and persistence of majority positions so that the mean opinion can shift even for a symmetric network. The model also formulates a distinction between opinion and rhetoric in accordance with a recently proposed theory of the group polarization effect. This enables the modeling of discussion-induced shifts toward the extreme without the typical modeling assumption of greater resistance to persuasion among extremists. An experiment is described in which triads engaged in online discussion. Simulations show that the ASC model is in qualitative and quantitative agreement with the experimental data.}

\section{Introduction}

While the experimental study of social influence and opinion change in particular primarily remains the province of the social sciences, the modeling of social influence dynamics, however, has extended into other fields including physics, computer science, and electrical engineering \cite{CasForLor2009,Kempeetal2016,ProTem2017}. The primary goal of opinion network models is to predict final opinions from initial ones typically via a process that updates node opinions over time. Continuous opinion models --- the concern of this paper --- allow for incremental shifts in opinion where the amount of change depends upon the distance between node opinions and the network of interpersonal influence that couples nodes. The DeGroot and Friedkin-Johnsen models, as well as the consensus protocol (a continuous time version of the DeGroot model), use a linear dependence in which the shift is proportional to the opinion difference \cite{DeGroot1974,FriJoh2011,OlfFaxMur2007}. Bounded confidence models posit a hard opinion difference threshold, within which nodes interact linearly, but beyond which the interaction vanishes \cite{Lorenz2007}. The nonlinear model of \cite{Gabbay2007a} uses a soft threshold so that, rather than vanishing completely, the interaction decays smoothly with distance.

Modeling how opinions become more extreme has been of particular concern in the opinion network modeling literature. The primary contribution of this paper is to present an opinion network model, the Accept-Shift-Constrict (ASC) model, which provides an experimentally supported depiction of the group polarization effect, a classic social psychology effect in which discussion among like-minded group members tends to make groups more extreme. The ASC model describes opinion change processes over a network as group members exchange messages.  These processes consist of, first, the acceptance of a persuasive message which can then lead to a shift in the receiver's opinion and also a constriction of the receiver's uncertainty level. In turn, this constriction narrows the extent to which subsequent messages advocating distant opinions are accepted.

This paper proceeds as follows. The next section discusses the group polarization effect along with its treatment in social psychology and the opinion network modeling literature. Section~\ref{sec:Experiment} describes a recent experiment involving discussion about betting on National Football League (NFL) games, the results of which challenge existing group polarization theory. In Sec.~\ref{sec:Frame}, an alternative \emph{frame-induced} theory of group polarization is presented that can account for the experimental results. Sections~\ref{sec:asc} and \ref{sec:simresults} present the ASC model and experimentally-relevant simulation results.

\section{Group Polarization Effect} \label{sec:polarization}

In the group polarization effect, discussion among group members who are all on the same side of an issue induces more extreme decisions or opinions (``polarization'' as used here connotes a group shifting further toward one pole of an issue rather than diverging toward opposite poles as in conventional usage) \cite{MyersLamm1976,Isenberg1986,Sunstein2002}. It was originally referred to as the ``risky shift'' effect as it was discovered in an experimental context involving small groups faced with choosing among options of varying risk levels; discussion tended to shift groups toward riskier options than the average of their pre-discussion preferences. Subsequent research observed systematic discussion-induced extremism in homogeneous groups in broader contexts including social and political attitudes and the severity of punishments in jury deliberations. A group is considered to be homogeneous with respect to an issue if all its members have initial preferences that lie on one side of the issue's neutral reference point. Group polarization is then said to occur if after the discussion the mean preference of the group shifts further away from the reference point compared with the mean prior to discussion. Polarization is typically observed for issues that have a substantial judgmental component as opposed to issues like math problems that have demonstrably correct solutions.

Two distinct processes, based on informational and normative influence respectively, are most commonly accepted in social psychology as causes of group polarization \cite{MyersLamm1976,Isenberg1986}. The informational influence explanation, known as persuasive arguments theory, focuses on the role of novel arguments. In essence, members of a homogeneous group, although inclined toward the same side of an issue, will typically possess different arguments in support of that side. The exchange of these arguments in discussion then exposes group members to even more information supporting their initial inclination and so shifts it further in the same direction. The normative influence explanation, social comparison theory, posits that the relationship of group member positions with respect to a culturally salient norm is critical rather than the information underlying those positions. The norm is taken to favor one pole of the issue. For example, a norm favoring risk-taking makes riskier positions more socially ideal than cautious ones. A major problem of the informational and normative influence theories is that they always predict polarization for an individual group whenever the polarization preconditions (homogeneous group and judgmental issue) are present, regardless of the distribution of initial opinions within the group. This problem stems from the fact that these theories were never reconciled with stronger, concurrent social influence phenomena such as majority influence and consensus pressure \cite{Gabbayetal2018}.

Within the opinion network modeling literature, extremism has been predominantly modeled by attributing higher network weights to nodes with more extreme initial opinions \cite{DefAmbWei2002,Friedkin2015}.  This approach, which we refer to as ``extremist-tilting,'' is necessitated by the property of most continuous opinion models that the mean opinion in networks with symmetric coupling remains constant at its initial value --- a property that is at odds with the shift in mean exhibited in group polarization. Consequently, extremists must be assigned greater influence over moderates than vice versa in order to shift the mean. This explanation is different from the two more prominent theories above but shares their problem of uniformly predicting polarization for homogeneous groups.


\section{Experiment} \label{sec:Experiment}

This section describes the group polarization experiment conducted in \cite{Gabbayetal2018} in which three-person groups engaged in online discussion about wagering on National Football League (NFL) games. As is standard practice in NFL betting, spread betting was employed rather than wagering directly on which team will win the game. In spread betting, the terms ``favorite'' and ``underdog'' refer, respectively, to the likely winner and loser of the game itself. The point spread is the expected margin of victory of the favorite team as set by Las Vegas oddsmakers. A bet on the favorite is successful if its margin of victory exceeds the spread; otherwise a bet on the underdog is successful.\footnote{In actual practice, bets are returned if the victory margin equals the spread.} If Team A is the favorite by a spread of six points over the underdog Team B, then Team A has to win the game by more than six points in order for a bet on Team A to pay off. The objective of the spread is to endeavor to equalize the odds for either the favorite or underdog to win the bet.

In the experiment, an upcoming NFL game was chosen and a pre-survey then elicited subject initial preferences with respect to team choice and a wager amount on that team from \$0 to \$7 (in whole-dollar increments). On the basis of the pre-survey, discussion groups were constructed with respect to three dichotomous variables. The first is \emph{policy side} of favorite or underdog corresponding to the team chosen as more likely to beat the spread. This variable imposes the polarization precondition of having like-minded group members as the groups are homogeneous with respect to the fundamental policy question of which team will win the bet. The second variable is \emph{disagreement level} of ``high'' or ``low'' that depends upon the difference between the minimum and maximum wagers in the group. Each group consisted of low, intermediate, and high wager individuals with respective wagers $w_1$, $w_2$, and $w_3$. In all groups, the intermediate wager was set so that $w_2 \in \{\$3,\$4\}$. In the high disagreement condition, $w_1=\$0$ and $w_3=\$7$ giving a difference of \$7. In the low disagreement condition $w_1 \in \{\$1,\$2\}$ and $w_3 \in \{\$5,\$6\}$ so that the difference could be \$3, \$4, or \$5. The third variable is \emph{network structure} of ``complete'' in which all members could communicate with each other or ``chain'' in which the intermediate wager member $w_2$ served as the center node connecting $w_1$ and $w_3$.  After discussion, each member made their final wager. A group decision was not required but groups arrived at a consensus wager far more often than the alternative outcomes of a two-person majority or three different wagers. A winning (losing) bet resulted in a payoff of \$7 plus (minus) the wager, which was donated to a charity.

\begin{figure}
\vspace{0.5in}
\begin{minipage}{0.5\textwidth}
\centering
\includegraphics[scale=0.4]{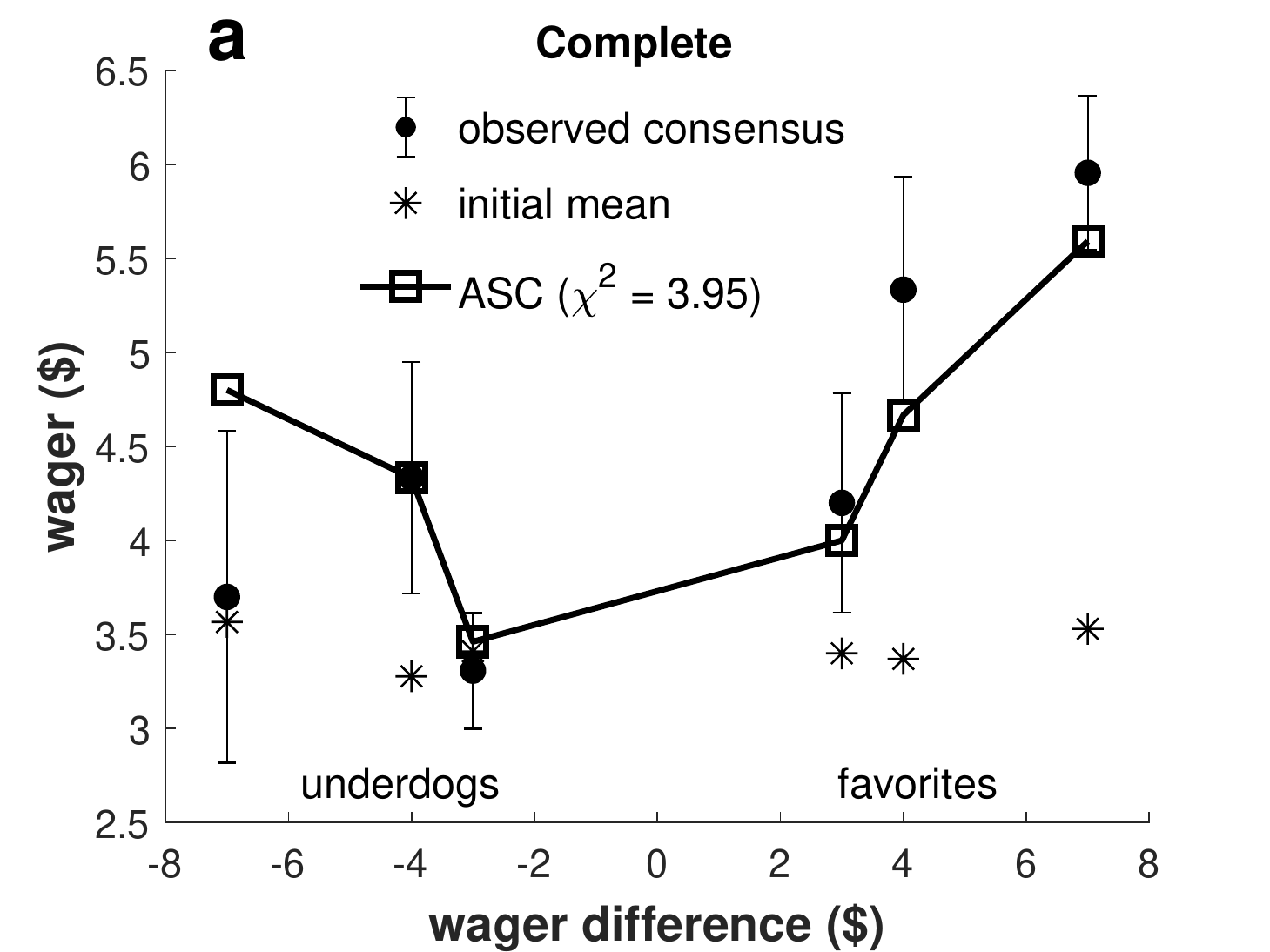}
\end{minipage}
\begin{minipage}{0.5\textwidth}
\centering
\includegraphics[scale=0.4]{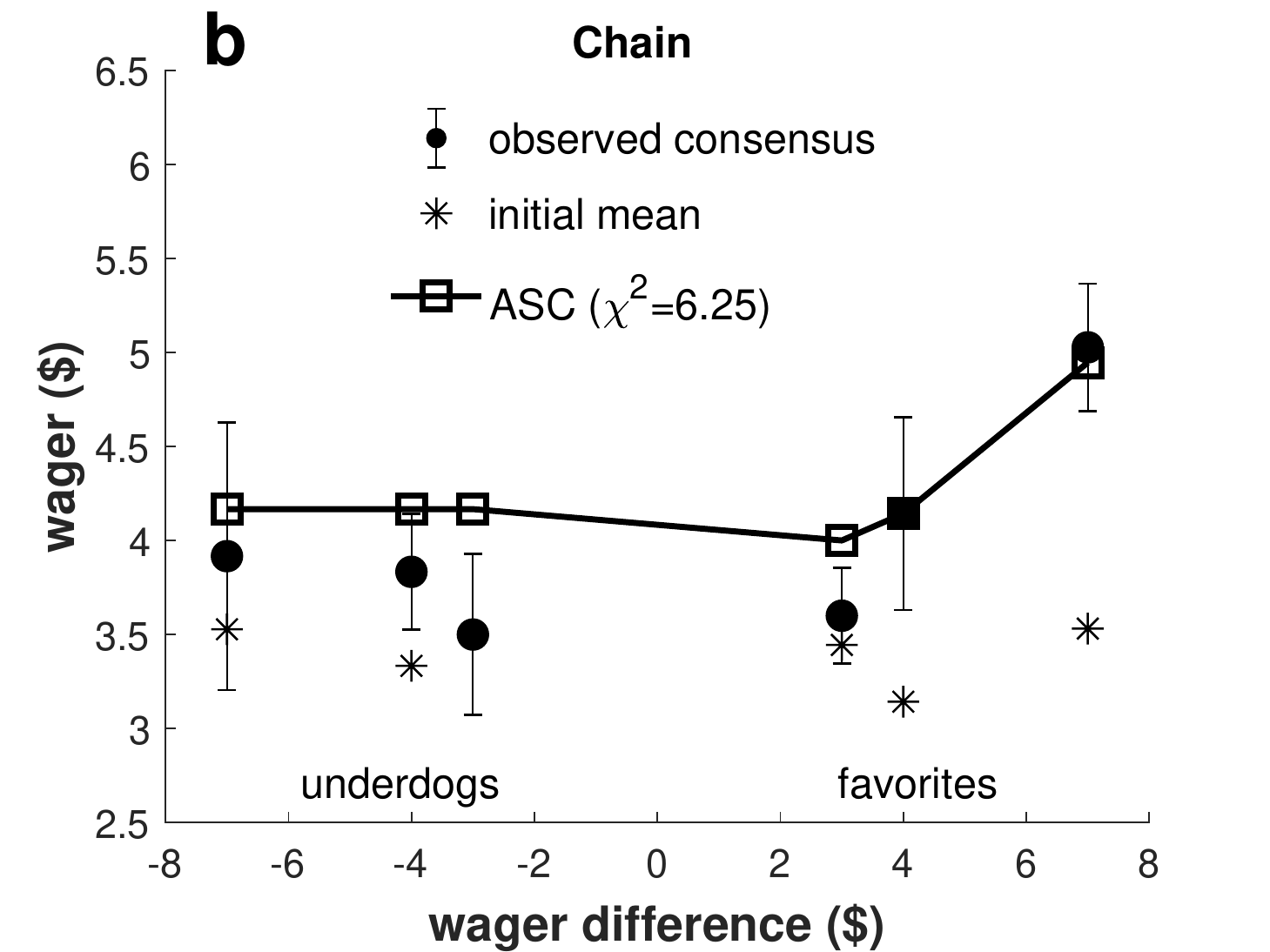}
\end{minipage}
\caption{Observed and simulated mean consensus wagers as function of initial wager difference, $w_3-w_1$. (\textbf{a}) Complete network. (\textbf{b}) Chain network.  Favorite groups shown on the positive $x$ axis, underdogs on the negative. Observed consensus is average of final consensus wager (taken as positive for both favorites and underdogs) over groups at each difference value (no \$5 difference groups were used as there were only four total). Also shown is average of the group mean initial wager.  Experimental data shown as circles. Error bars are standard errors. $\chi^2$ value is the sum of the squared errors between the simulated and the experimental values normalized by the standard error at each data point. Simulation parameters: $\alpha=0.034, \lambda(0)=0.03, \lambda_{min}=0.01$.}
\label{fig:expplot}
\end{figure}

Polarization, or more specifically a risky shift, is observed for a group if its mean wager after discussion is greater than its initial mean wager. Most of the 198 groups reached a consensus wager. For these 169 consensus groups, statistically significant results were observed for all three of the manipulated variables. For policy side, only the favorite side exhibited a risky shift whereas the underdog side did not. For disagreement level, restricted to favorite groups (as underdog groups showed no systematic risky shift), high disagreement groups exhibited a greater risky shift than low disagreement groups. For network structure, similarly restricted to favorites, complete networks showed a greater risky shift than chains. All three of these behaviors can be seen in Fig.~\ref{fig:expplot} in which substantial polarization is observed when the error interval is above the initial mean.

The above results are not readily explained by standard polarization theory. Particularly challenging is the policy side result as standard theory predicts that both policy sides should show a risky shift. For persuasive arguments theory, members of both the favorite and underdog groups presumably possess novel information in support of their team choice and should therefore increase their confidence and wager. For social comparison theory, a norm toward risk taking should cause both sides to increase their wager. The extremist-tilting explanation prevalent in opinion network modeling also fails to explain this differential polarization by policy side: if individuals with more extreme wagers are taken to be more confident and persuasive, then both favorite and underdog groups should display an equal tendency to increase their wagers.

\section{Frame-Induced Polarization Theory} \label{sec:Frame}
Reference~\cite{Gabbayetal2018} proposes a novel theoretical mechanism for group polarization that explains the results of the experiment. Central to the proposed mechanism is the distinction between the quantitative policy under debate and the \emph{rhetorical frame} --- the aspect of the policy upon which deliberations focus. The rhetorical frame will typically correspond to the dominant source of disagreement within the group due, for instance, to uncertainty as to the likelihood of an outcome. In a binary gamble such as in the experiment, the policy (e.g. wager amount) on a given outcome (e.g. team) and the rhetorical frame should be the subjective probability that that outcome will occur (e.g. win against the spread). The rhetorical frame position $\rho(x)$ is taken to be a function of the policy $x$. Groups will tend to shift toward the extreme if the functional relationship between the rhetorical position and the policy is concave ($\rho^{\prime\prime}<0$), that is, the rhetorical position increases more slowly as the policy becomes more extreme. For the experiment, such a concave relationship is expected between the subjective probability that a subject's chosen team will win against the spread and the wager amount (see Sec.~\ref{sec:simresults}).

\begin{figure}
\centerline{\includegraphics[width=.7\textwidth]{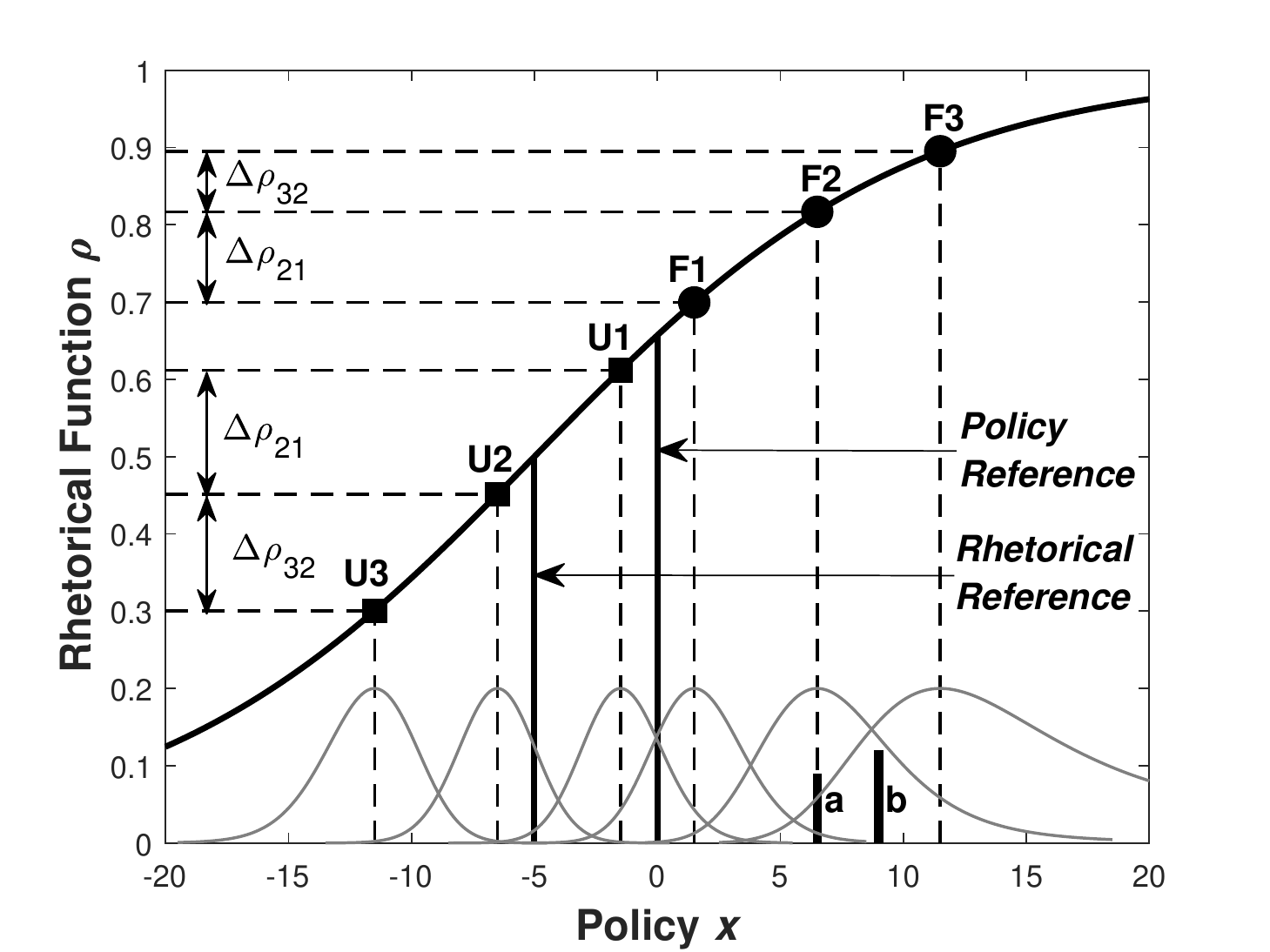}}
\caption{Effect of rhetorical function concavity and offset reference.$\rho(x)=1/(1+e^{-\beta(x-x_0)})$ with $\beta=0.13, x_0=-5$. Short lines at bottom show alternative $F$ group consensus policies: (a) mean policy $\bar{x}=x_2$; (b) RPM policy $\bar{x}_{23}=(x_2+x_3)/2$. ASC model acceptance functions in gray.}
\label{fig:riacurve}
\end{figure}

The effect of concavity is to compress rhetorical distances toward the extreme relative to the distances between more moderate members, making it easier for majorities to form on the extreme side of the mean. Consequently, while the policy distribution may be symmetric so that no majority is favored on either side of the mean (as is approximately the case in our experiment), the distribution of rhetorical positions is skewed so that there is an initial majority on the extreme side of the rhetorical mean. This rhetorically-proximate majority (RPM) converges to a policy position more extreme than the mean to which the moderate minority of group members (those with policies below the mean) then concur, thereby resulting in a consensus policy that exhibits group polarization. The members of the $F$ group (analogous to the favorite groups) in Fig.~\ref{fig:riacurve} provide an example of this mechanism. Although the intermediate member $F_2$ is equidistant in policy from the moderate $F_1$ and the extremist $F_3$, $F_2$ is rhetorically closer to $F_3$ and therefore $(F_2,F_3)$ is the RPM pair. They agree on a policy halfway between them to which $F_1$ comes up due to majority influence. The RPM policy (b) is seen to be greater than the initial mean policy (a).

Although the concavity of the rhetorical function explains the basic group polarization effect, it cannot by itself account for unequal polarization on opposing policy sides as observed in the experiment. Capturing this differential polarization involves the freedom of the rhetorical function to have a different reference point than the policy. The policy reference is defined as the neutral point, taken to be $x=0$, that demarcates opposing policy sides. The rhetorical reference is defined as the policy value that maps to the neutral point of the rhetorical frame. For a \emph{proper} frame, the rhetorical reference is the same as the policy reference so that the pro and con policy sides coincide with the pro and con rhetorical sides. For an \emph{improper} frame, the rhetorical and policy references are offset so that the rhetorical reference splits one of the policy sides. Figure~\ref{fig:riacurve} shows how an improper frame can lead to differential polarization by policy side. The rhetorical reference splits the con (negative) policy side, which results in the $U$ group (analogous to underdog groups) being arrayed on the approximately linear part of the rhetorical function rather than on the shoulder as for the $F$ group. Consequently, $U_2$ is roughly the same rhetorical distance from both $U_1$ and $U_3$. Considering the effects of uncertainty and noise, formation of the moderate $(U_1,U_2)$ RPM pair is about as likely as the extreme $(U_2,U_3)$ pair so that systematic group polarization is absent or much reduced as observed in the experiment for the underdog groups. An improper frame can result from the heuristic substitution of a simpler, intuitive frame in place of a more complex proper frame that directly corresponds to the policy \cite{Gabbayetal2018}. In the experiment, the heuristic frame of which team will win the \emph{game} replaces the proper frame of who will win against the \emph{spread}.

\section{Accept-Shift-Constrict Model} \label{sec:asc}

The ASC model evolves both the positions and uncertainties of group members in response to their dyadic interactions. We consider position first, which can be a policy or, more generally, an opinion about some matter. A persuasive message sent by one group member to another must first be accepted by the recipient in order to shift their policy. While a number of factors can affect whether a message is accepted, the distance between the message's rhetorical position and that of the receiver plays the key role in our model: if the distance is within the \emph{latitude of acceptance} (LOA), the message is likely to be accepted, but the acceptance probability rapidly decays beyond the LOA. If the message is accepted, then the receiver's policy is shifted in proportion to its distance from the sender's policy.

Formally, we encode the above process as an ordinary differential equation for $x_i(t)$, the policy position of the $i^{th}$ group member at time $t$.  For a group with $N$ members, the rate of change of $x_i$ is given by
\begin{equation}
\frac{dx_i}{dt} = \sum_{j=1}^{N} \nu_{ij} (x_j-x_i) \exp{\left\{ -\frac{1}{2}
\frac{(\rho(x_j)-\rho(x_i))^2}{\lambda_i^{2}} \right\}},
\label{eq:dxdt}
\end{equation}
where $\nu_{ij}$ is the coupling strength from $j \rightarrow i$ and $\lambda_i$ is $i$'s LOA. The matrix formed by the coupling strengths defines a position-independent network of influence. In general, $\nu_{ij}$ depends on communication rate and other factors such as credibility and expertise ($\nu_{ii}=0$).

The linear $x_j-x_i$ term in Eq.~(\ref{eq:dxdt}) represents the shift effect. The gaussian term represents the acceptance process and we refer to it as the acceptance function, $a(\Delta\rho,\lambda)=e^{-\Delta\rho^2/2\lambda^2}$. Although the acceptance function is always symmetric with respect to the sign of the rhetorical difference, $a(-\Delta\rho)=a(\Delta\rho)$, a concave $\rho(x)$ can causes it to appear asymmetric along the policy axis as clearly seen for $F_2$ and $F_3$ in Fig.~\ref{fig:riacurve}.

In addition to position change, communication can also affect a person's uncertainty regarding their position. Group discussion has been observed to increase the level of certainty that members have in their quantitative judgments \cite{Sniezek1992}. Accordingly, we introduce an uncertainty reduction mechanism in our model in which messages from those with similar positions constrict an individual's LOA so that they become more resistant to persuasion from distant positions. Messages originating within the LOA that are accepted decrease the LOA, but not beneath a certain minimum value $\lambda_{min}$. This yields for the LOA dynamics:
\begin{equation}
\frac{d\lambda_i}{dt}=\left\{
\begin{array}{ll}
     \sum_{j=1}^{N} \nu_{ij} (\lambda_{min}-\lambda_i) e^{ -\Delta\rho_{ij}^2/2\lambda_i^{2}},
     & |\Delta\rho_{ij}| \leq \lambda_i \\
     0, & |\Delta\rho_{ij}| > \lambda_i.
\end{array}
     \right.
\label{eq:loa}
\end{equation}

Equations~(\ref{eq:dxdt}) and (\ref{eq:loa}) comprise the ASC model. Assuming no difference between rhetorical and policy positions, i.e, $\rho(x)=x$,  Eq.~(\ref{eq:dxdt}) is equivalent to the model of \cite{Gabbay2007a} without the self-influence force that models a persistent effect of an individual's initial opinion. The uncertainty reduction dynamics represented by Eq.~(\ref{eq:loa}) is novel in opinion network modeling. The model of \cite{DefAmbWei2002} includes a dyadic uncertainty interaction that results in uncertainty change only when dyad members have different uncertainties; this requires that uncertainty levels be visible to other group members, an assumption not present in Eq.~(\ref{eq:loa}), and does not allow equally uncertain individuals to mutually reinforce their opinions.

\begin{figure}
\vspace{1in}
\begin{minipage}{0.5\textwidth}
\centering
\includegraphics[scale=0.4]{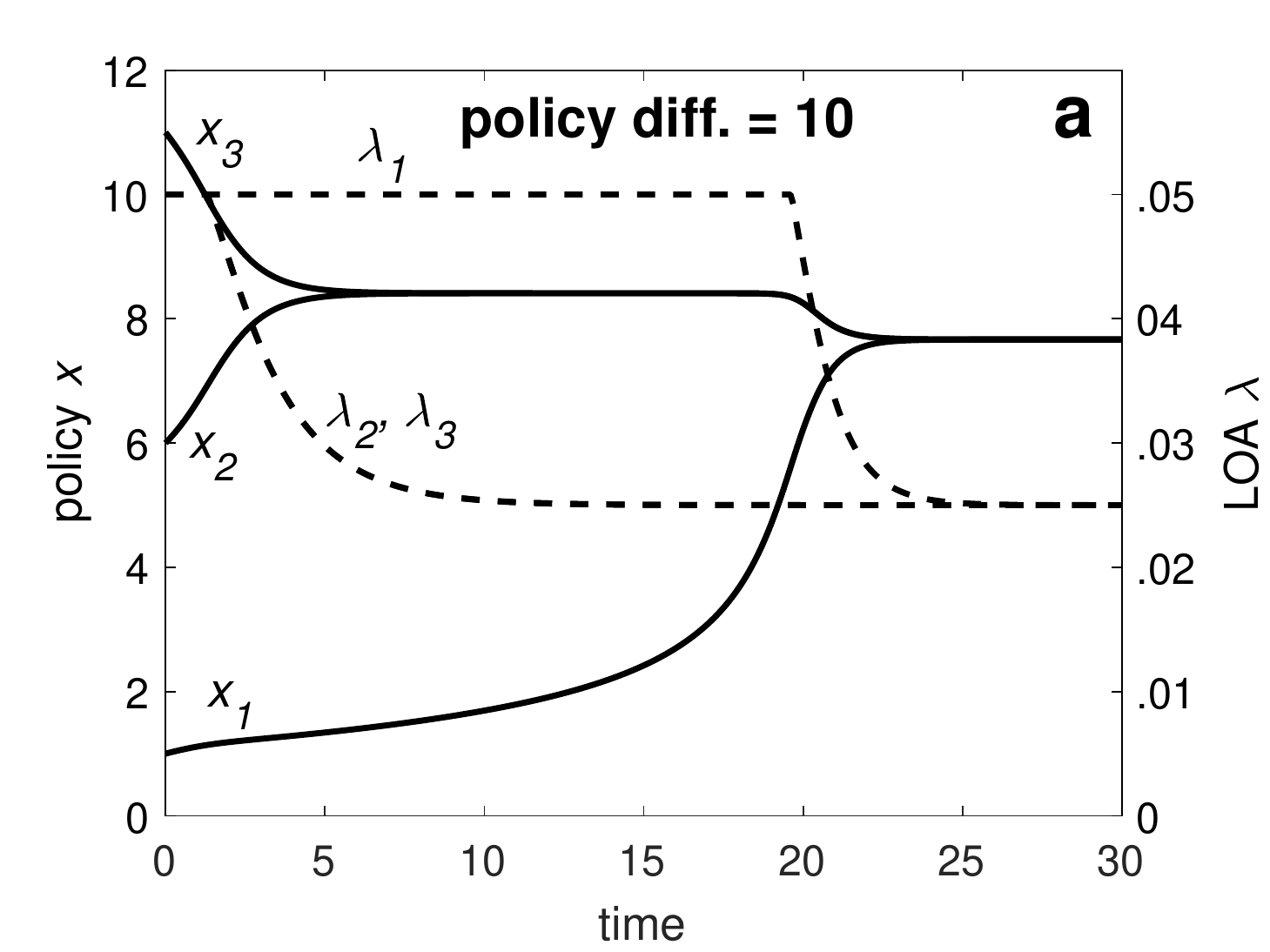}
\end{minipage}
\begin{minipage}{0.5\textwidth}
\centering
\includegraphics[scale=0.4]{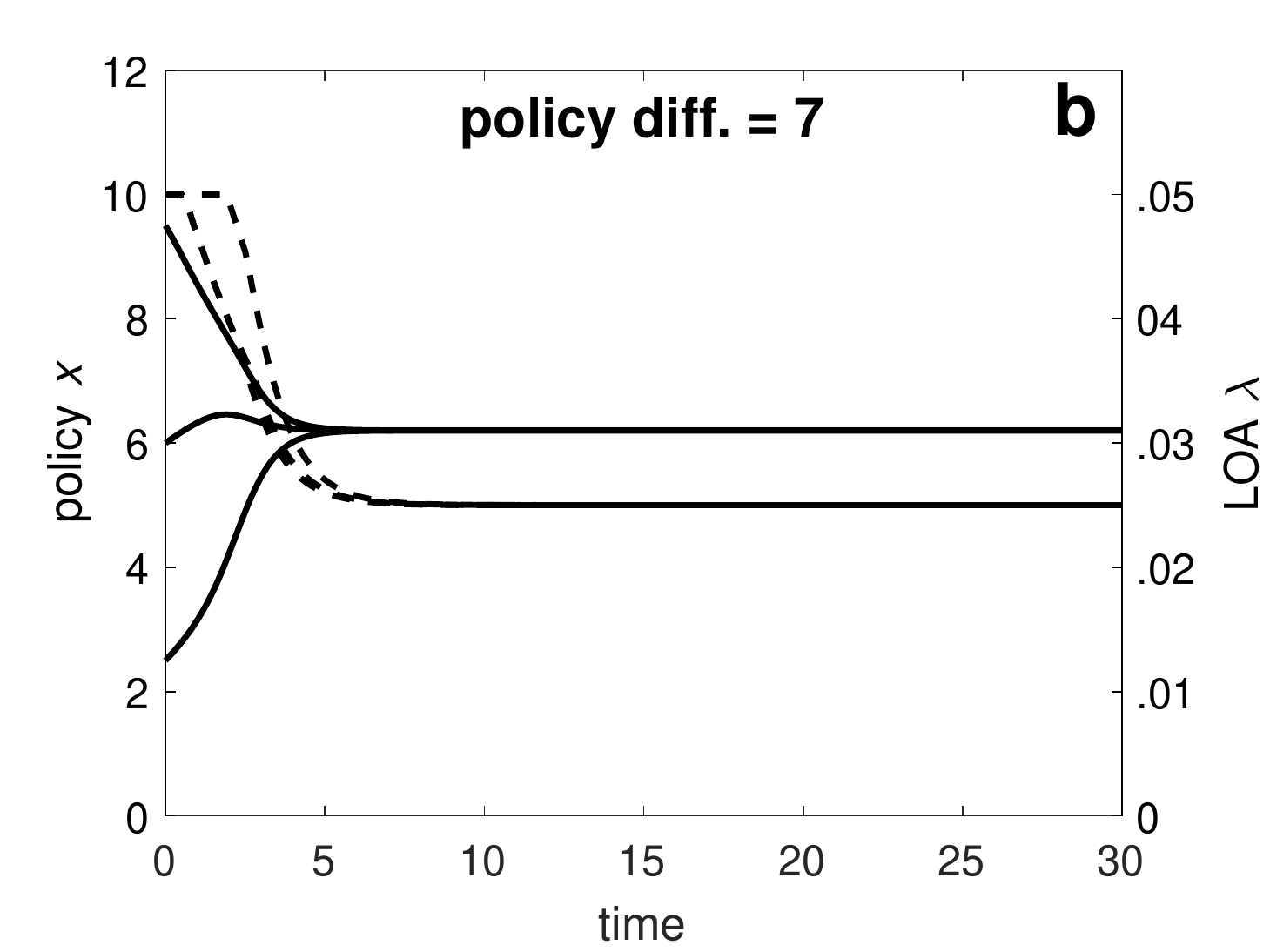}
\end{minipage}
\caption{Position and LOA trajectories in ASC model for a complete network. Solid curves show policy positions, dashed curves show LOAs. (\textbf{a}) High initial policy disagreement ($x_3-x_1=10$) showing substantial shift between consensus and initial mean policy ($x_2(0)$). (\textbf{b}) Lower initial policy disagreement (7) results in near simultaneous convergence close to initial policy mean.  $\lambda_{1,2,3}(0)=0.05, \lambda_{min}=0.025$; $\rho(x)$ as in Fig.~\ref{fig:riacurve}.}
\label{fig:asctraj}
\end{figure}

A crucial consequence of the uncertainty reduction dynamics in the ASC model is the ability for interim majorities to more effectively maintain their position in the face of minority influence. This effect is essential to the RPM process in the theoretical account of group polarization above but it occurs regardless of whether or not the rhetorical function is different from the policy. Figure~\ref{fig:asctraj}a illustrates the rough persistence of the majority position for a complete-network triad in which the intermediate member's position is taken to be halfway between the others. For sufficiently low initial disagreement, however, an interim majority will not form and the group equilibrium will be close to its initial mean (Fig.~\ref{fig:asctraj}b).\footnote{The persistence of majority positions on a continuous opinion axis is also found in the agent-based model of \cite{Mouetal2013}, which employs a confidence variable that must be transmitted between agents along with opinions, rather than the ASC model's use of an uncertainty interval not visible to others.}

\section{Simulation of Group Polarization} \label{sec:simresults}

This section demonstrates the ability of the ASC model to produce the same qualitative effects as in the frame-induced polarization theory and as observed experimentally. Going beyond qualitative correspondence, its agreement with the data on a quantitative level is also shown. First, we discuss how the coupling strengths $\nu_{ij}$ are set. They are treated as dyadic communication rates as determined by simple topological considerations. For a complete network, on average, the communication rates are expected to be the same for all nodes, so we set $\nu_{ij}=1/2$ for all three dyads.  For the chain, if the sequence in which nodes send messages follows the chain path and the center node (node 2) predominantly opts to send its messages simultaneously to both outer nodes (rather than separately), then we expect node 2 to have about twice the communication rate with each of nodes 1 and 3. We therefore set $\nu_{12}=\nu_{32}=1$ and $\nu_{21}=\nu_{23}=1/2$.\footnote{The sum of the communication weights is normalized to the same (arbitrary) value of 3 in both networks, a value that only affects the transient time and not the final equilibrium.} These communication rate expectations are approximately borne out in our experiment.

\begin{figure}

\begin{minipage}{0.5\textwidth}
\centering
\includegraphics[scale=0.35]{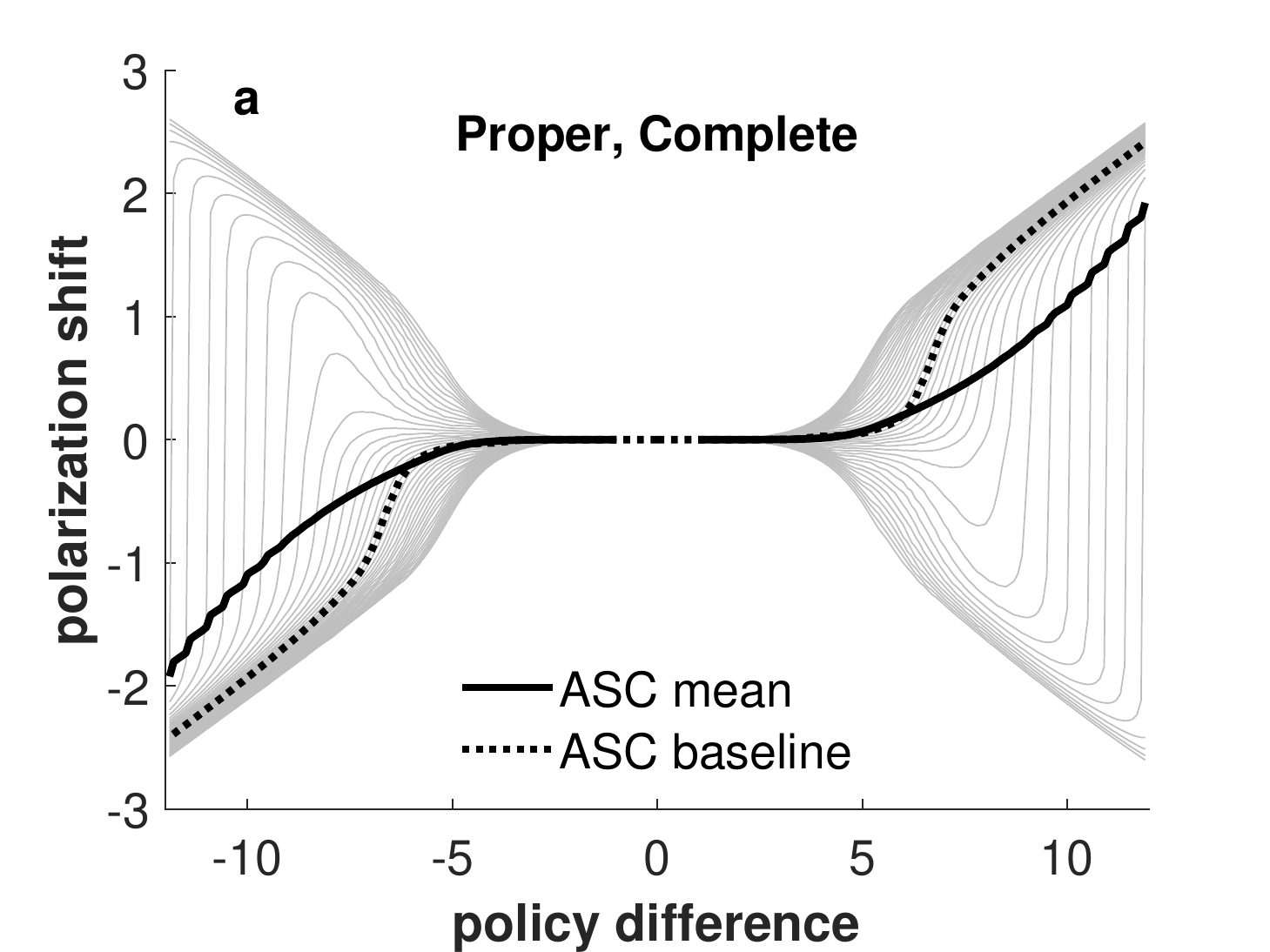}
\end{minipage}
\begin{minipage}{0.5\textwidth}
\centering
\includegraphics[scale=0.35]{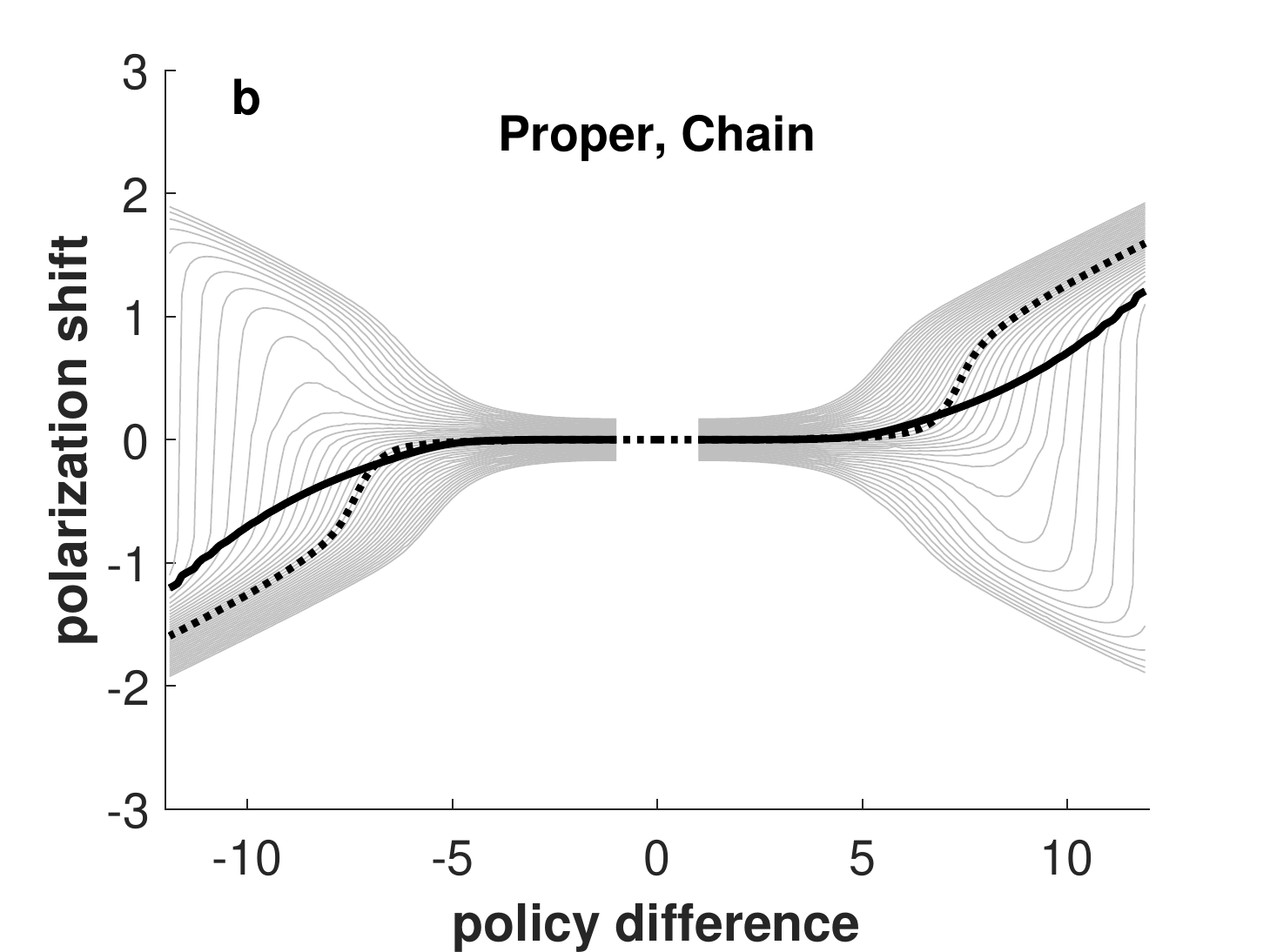}
\end{minipage}
\begin{minipage}{0.5\textwidth}
\centering
\includegraphics[scale=0.35]{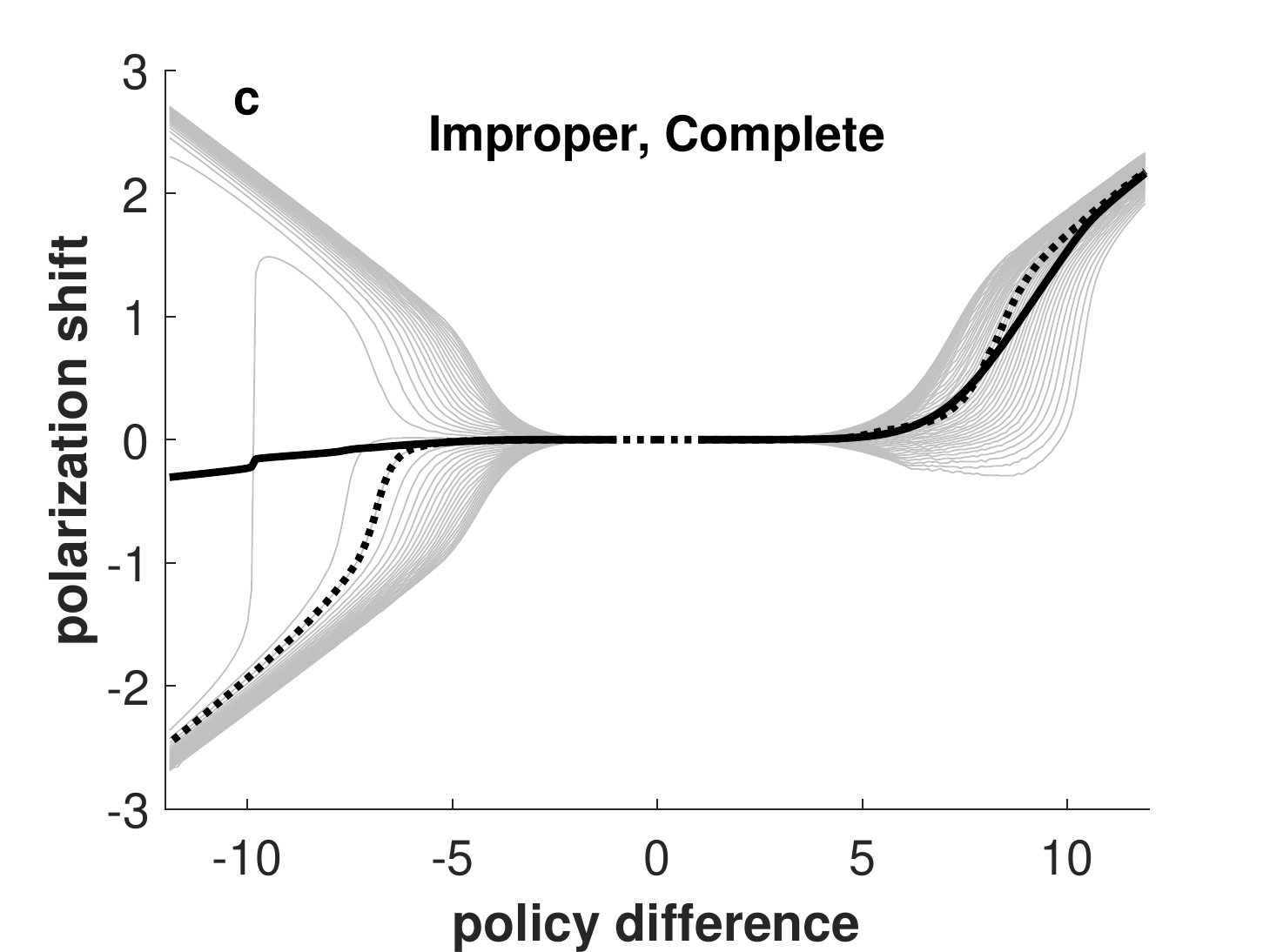}
\end{minipage}
\begin{minipage}{0.5\textwidth}
\centering
\includegraphics[scale=0.35]{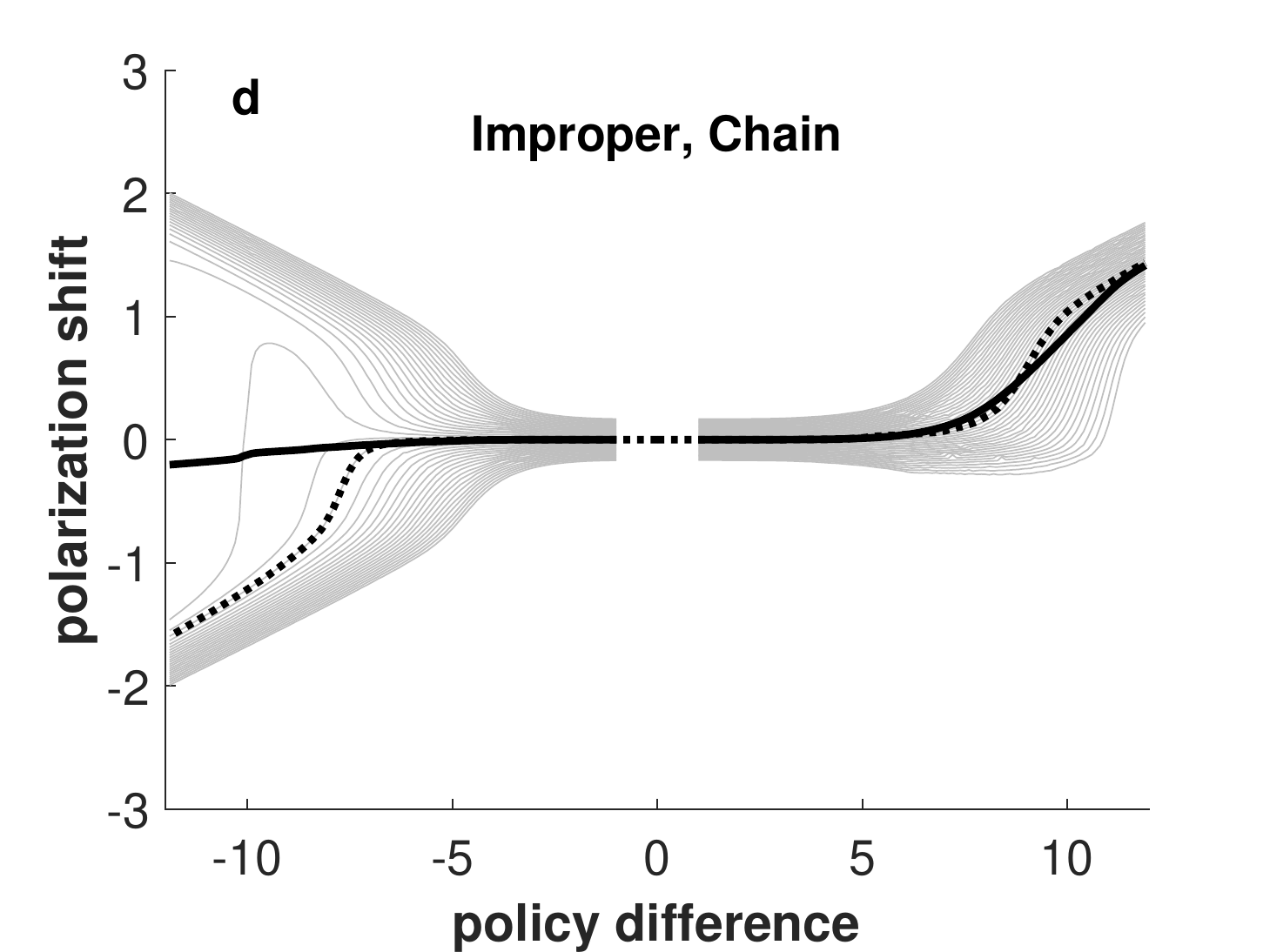}
\end{minipage}
\caption{ASC simulations for triad networks with variability in intermediate node policy.  $\rho(x)$ taken as in Fig.~\ref{fig:riacurve}. Top row shows proper rhetorical frame ($x_0=0$).  Bottom row shows improper rhetorical frame ($x_0=-5$). Positive and negative policy sides are on positive and negative horizontal axis respectively. Polarization shift, $\delta=\bar{x}(t_f)-\bar{x}(0)$, plotted as a function of the initial policy difference, $\Delta=x_3(0)-x_1(0)$. Shift toward the extreme corresponds to $\delta > 0$  for positive policy side and $\delta<0$ for negative side. The position of the intermediate node was varied according to $x_2(0)=\pm (6 +\epsilon)$ for the positive and negative policy sides, where $\epsilon$ takes on 41 uniformly-spaced values over the interval $[-1,1]$.  $x_1(0)=6-\Delta/2$ and $x_3(0)=6+\Delta/2$ for $\Delta>0$ and analogously for $\Delta<0$. ASC mean (black) taken over all $\epsilon$ values. Shifts for individual $\epsilon$ values shown as gray curves. Dotted curve shows $\epsilon=0$ baseline case where $x_2(0)=6$. Gap in the curves is the region where $x_2(0)$ would go beyond $x_1(0)$ or $x_3(0)$. ASC model parameters: $\lambda_{1,2,3}(0)=0.05, \lambda_{min}=0.025.$ }
\label{fig:ascsims}
\end{figure}

Figure~\ref{fig:ascsims} displays simulation results for complete and chain network triads that are homogeneous with respect to policy side analogous to the experimental setup. The baseline case (dotted curve) consists of an intermediate node with an initial policy $x_2(0)$ halfway between the initial positions of the moderate $x_1(0)$ and the extremist $x_3(0)$. The other cases shown (light gray curves) account for position uncertainty by allowing $x_2(0)$ to deviate by various small amounts from the baseline case. The discussion-induced shift in the mean is plotted against the initial policy difference between the extremist and the moderate, where the opposing pro and con policy sides are shown on the positive and negative sides of the horizontal axis respectively. For the pro (con) side, a positive (negative) polarization shift indicates a shift toward the extreme --- a higher wager in the case of the experiment. The mean over all the cases (solid dark curve) can be used to gauge the extent of systematic polarization.

The top row of Fig.~\ref{fig:ascsims} represents a proper rhetorical frame in which the policy and rhetorical references are coincident. In the experiment, the proper frame is the subjective probability of the favorite winning against the spread. The rhetorical function is taken to be concave with increasing policy extremity.\footnote{If the subjective probability of one of the binary outcomes is taken as the rhetorical frame and opposing policy sides have opposite signs, then concavity with increasing policy extremity yields an overall S-shaped rhetorical function as explained in \cite{Gabbayetal2018}.} Regarding the mean, both policy sides exhibit equal polarization that increases with disagreement level and with the complete network showing more polarization than the chain. Considering higher disagreement levels, the mean polarizes less than the baseline case because some groups actually depolarize --- those in which the moderate and intermediate node are sufficiently close to overcome the skewing effect of the rhetorical function. This ability to predict depolarization for individual groups despite the dominant tendency toward polarization is an important capability not present in the informational, normative, or extremist-tilting theories. Although the proper frame does exhibit polarization, accounting for the differential polarization by policy side observed experimentally requires use of an improper frame as is the subjective probability that the favorite will win the game. The bottom row of Fig.~\ref{fig:ascsims} employs an improper frame and indeed shows substantial polarization for positive policies and little for negative ones.

The ASC model can also be quantitatively tested against the data. Groups can be simulated using their actual initial wagers and with the communication weights as set above. The rhetorical function $\rho(w)$ that maps the wager to the subjective probability of a favorite game victory (the improper frame) is derived in \cite{Gabbayetal2018} based on the theory of individual decision making under risk and uncertainty. It depends upon the subjective probability $p(w)$ of a favorite victory (the proper frame)
\begin{equation}
p(w) = \frac{1}{2} - \frac{1}{8\alpha w} \pm \frac{1}{2} \sqrt{1 + \frac{1}{16 \alpha^2 w^2}},
\label{eq:pw}
\end{equation}
where the + ($-$) sign implies bets on the favorite (underdog). The free parameter $\alpha$ is the risk aversion that quantifies how sensitive individuals are to variance around the expected value of the payoff. It is assumed to be identical for all subjects. The rhetorical function is then given by
\begin{equation}
\rho(w) = \frac{1}{2} \mathrm{erfc} \left\{ \mathrm{erfc}^{-1}\left(2 p(w)\right) - \frac{s_0}{\sigma \sqrt{2}} \right\},
\label{eq:rho}
\end{equation}
where $\mathrm{erfc}(u)=\frac{2}{\sqrt{\pi}}\int_{u}^{\infty} e^{-v^2}dv$. The parameter $s_0$ is the point spread for the game in question and $\sigma=12.8$ is the empirical standard deviation for the margin of victory in NFL games. Both $p(w)$ and $\rho(w)$ are S-shaped implying a concave relationship between the subjective probability of the outcome estimated as more likely and the wager magnitude.

In addition to the risk aversion, there are two free parameters from the ASC model that need to be fit from the data, the initial LOA $\lambda(0)$ and the minimum LOA $\lambda_{min}$, both assumed identical for all subjects. The three parameters are estimated by minimizing the sum of $\chi^2$ error values over both complete and chain networks. The simulation results are shown in Fig.~\ref{fig:expplot}. A three-parameter $\chi^2$ goodness-of-fit test, which takes as its null hypothesis that the model is correct, yields a probability $Q=0.33$ that $\chi^2$ could have exceeded its observed value of 10.2 by chance. With a conservative threshold of $Q < 0.2$ for rejecting the null hypothesis, the ASC model is found to be consistent with the data.

\section{Conclusion} \label{sec:Conclusion}

The ASC model presented here describes a dual process of opinion and uncertainty change based on the greater acceptance rate of messages within one's LOA and the decrease in LOA due to exposure to similar views. A key dynamic in the model is the ability of proximate majorities to form and persist for symmetric networks, thereby enabling majorities to exert outsized influence and produce a consensus opinion different from the initial mean. Importantly, the ASC model does not involve the exchange of uncertainties over the network unlike other models in which uncertainties are directly coupled along with opinions \cite{DefAmbWei2002,Mouetal2013}.  Another important innovation of the ASC model is the conceptualization of distinct dimensions of opinion and rhetoric: opinion is an evaluation directly tied to a decision or other behavioral outcome of interest while rhetoric determines whether messages aimed at shifting opinions are found persuasive. If the rhetorical function mapping opinion to rhetorical position is concave, then proximate majority formation at the extreme is facilitated. Consequently, the ASC model can generate systematic group polarization due to the structure of the decision space rather than by assuming an asymmetric network structure in which influence is associated with extremity as typically done in opinion network modeling. The ASC model simulations shown here display the same qualitative phenomena as observed in the experiment: polarization on one policy side but not the other, increasing polarization with disagreement level, and greater polarization for complete networks than for chains. Furthermore, the ASC model is in quantitative agreement with the experimental data.


\begin{acknowledgement}
This work was supported by the Office of Naval Research under grant N00014--15--1--2549.
\end{acknowledgement}
%
%
%

\end{document}